\documentclass[12pt,twoside]{article}
\usepackage{xrb2007}
\usepackage{graphicx}
\begin{document}

\session{ULXs}

\shortauthor{Pakull}
\shorttitle{ULX Beambags}

\title{Ultraluminous X-ray Sources: Beambags and Optical Counterparts}
\author{Manfred W. Pakull \& Fabien Gris\'e}
\affil{Observatoire Astronomique, 11 rue de l'Universit\'e, 
F67000 Strasbourg, France}

\begin{abstract} 
A significant fraction of ultraluminous
X-ray  sources appear to be embedded in observable 
ionized nebul\ae\ that take the form of large, several
100\,pc diameter  interstellar bubbles.  Here we review
optical observations of these bubbles, their importance
for our understanding of the nature of ULXs,  the
energetics involved and their formation and evolution. 
Among the results obtained are new arguments against
conventional  superbubble scenarios and hypernova
remnants, and we present the case in favour of 
ULX-wind/jet
driven bubbles. We report the discovery of new ULXs in
very large SNR candidates in nearby galaxies, and finally
present an image of a triple X-ray source coincident with 
the radio-bright bubble S26 in the galaxy NGC~7793 which
appears to be a   clone of the microquasar SS433/W50
system.  
\end{abstract}

\section{Introduction} 
In recent years it has become obvious that the nature of
UltraLuminous  X-ray sources (ULXs) in nearby galaxies
cannot be elucidated from X-ray observations alone. In
particular, the question whether or not  ULXs harbour
specimen of the well-advertised intermediate mass black 
holes (IMBH) lying in between the stellar and the AGN
variety has stimulated  numerous investigations. Other
ideas put forward to overcome  the classical Eddington
luminosity limit  ($10^{39.1} \;
\mathrm{M}/(10\mathrm{M}_{\odot})$ erg/s)  of a stellar mass black hole
with M$\sim10$M$_{\odot}$ include beamed emission into our
line of sight - either geometrically or 
relativistically - and highly Super-Eddington accretion 
rates, presumably coupled with the occurrence of massive
outflows from the systems.

Optical follow-up observations of nearby ULXs have in many
cases revealed  the presence of extended ionized
nebul\ae\, \citep{pam02,pam03} and of stellar
associations \citep{grpm06,gr08} that both can be used to derive important 
constraints on possible formation scenarios, ages and on
the photon- and  mechanical luminosities involved. X-ray
photoionisation arguments have  previously been used to
exclude strongly-beamed X-ray emission from  the very
luminous ULX in Holmberg~II. In this contribution we will
mainly focus on the effects of mechanical power that is released
by ULXs and is injected into the ISM.

\section{Photoionisation versus Shock Excitation}
Galactic massive X-ray binaries are generally not located in
regions of  high interstellar density. That is why observable
X-ray photoionized  nebul\ae\ (XIN), as, e.g. N\,159F around the
black hole candidate LMC X-1  {\citep{pang86} are rare. The key
observation which indicated that X-ray  emission is largely
responsible for the high ionization structure of N\,159F came
from the detection of extended nebular He\,II$\lambda4686$
recombination radiation. This line is not present in normal
H\,II regions because even  the hottest O stars do not emit
significant amounts of He$^+$ Lyman continuum photons that
could ionize observable He\,III regions (exception:  hot
Wolf-Rayet stars with unusually weak winds).
An even more powerful XIN is embedded in the star forming
region  HSK~70 located in the M81 group dwarf galaxy
Holmberg\,II. This He\,III  region is excited by the bright
ULX in that galaxy \citep{pam02} and allows, through Zanstra-type 
$\lambda4686$ photon
counting arguments, to establish the flux of the ionizing
He$^+$ Lyman continuum to be compatible with the 
10$^{40}$ erg/s isotropic luminosity measured with X-ray
telescopes. In other words, optical observations have been employed to
exclude the possibility that the source is beamed towards us.
Possibly, a further XIN example is provided by the $\lambda4686$ 
emitting 30 pc diameter supernova remnant candidate MF16 around 
the ULX NGC~6946 X-1 \citep{abol}. However, in this case, the
observed  expansion velocity is quite high, $\sim230$ km/s
\citep{du00},  which implies that shock-ionization must play a
key role here.

We have previously commented on the similarity of optical spectra
displayed by XIN on the one hand and by shock excited gas on the
other hand.  The situation is analogous to the
ambiguity in the interpretation of spectra of the {\it LINER}
subgroup of AGN and of extended emission line regions {\it (EELR)}, where 
both shocks and ionisation by hard non-stellar continua have 
been put forward to explain their characteristics. 

\section{Structure and Energetics of Large ULX Bubbles}
 
It came as quite a surprise when it was realized that a
significant fraction  of ULXs are surrounded by large bubble-like
nebul\ae\ (hereafter ULXB) of  some 200-500 pc diameter. 
Their optical spectra are reminiscent of the
generally more than 10  times smaller supernova remnants,  i.e.
shock-ionized gas combining strong emission from low-ionisation
species  like [OI]$\lambda6300$ and [SII]$\lambda6716,31$ with
emission  from more highly ionized gas, as e.g.
[OIII]$\lambda5007$.
The latter forbidden transition is excited only at sufficiently high shock
speeds, v$_s$, and the diagnostic $\lambda5007$/H$\alpha$ ratio varies by a 
factor $\geq10$ in the very narrow critical range of 80-100 km/s 
(somewhat dependent on the pre-shock conditions) before leveling off at about 
unity at higher velocities \citep{dop84}.

Although the ULX counterparts often appear to be part of
moderately  rich stellar associations \citep{grpm06,gr08}, the
clusters  are mostly too old ($\geq$20-40 Myrs) to photoionize 
observable H\,II regions. Naturally, even v$_s\sim100$ km/s
shocks become unobservable once the IS density drops below, say,\,
$\sim0.3$ cm$^{-3}$, when  the H$\alpha$ recombination line
intensity behind the shock -- which is  proportional to the
preshock density $n$ (not \mbox{$\sim n^2$ !)} -- drops below a threshold
where fluctuations of the  diffuse galactic background emission
become important.   

Optical observations of ULX bubbles show an amazing variety of 
filamentary and shell morphologies, as well as varying nebular
excitation displayed by diagnostic emission line ratios. This is 
well illustrated by the huge, 500 pc diameter shell-like 
ULXB MH~9-11 around Holmberg~IX X-1 (Fig. 1) first described by \citet{mi95}, 
and which also displays expansion velocities in the critical range.  
The image underlines the markedly different OIII vs H$_{\alpha}$  morphology
in the bubble and shows a green-coded $\lambda5007$ dominated 
extension towards the South-East (MH\,11). A possible explanation is 
in terms of shocks
that propagate into a pre-ionized medium and which at the same time are
back-illuminated by the ionizing radiation of the ULX; see, e.g. \citet{du89}. 
This prevents the formation of a full recombination region, and the emission 
is characteristic of an uncomplete shock
\citep{ray88}.

Age and energetics of ULXB can readily be estimated from the
diameters and expansion velocities of these structures, depending
on whether we assume a SNR like nature or a continuously inflated
wind/jet-driven bubble, see, e.g. \citet{pagm06}. It turns out that
the 400 pc diameter, v$_s\sim100$ km/s, prototype ULXBs around
Holmberg~IX X-1 and NGC~1313 X-2 are typically $\sim1$ million
years old, have an energy content of \mbox{$\sim10^{53}$ erg}, or
equivalently, have experienced an average inflating wind/jet power of
\mbox{$\sim10^{40}$ erg/s} over  the age of the ULXB.



\begin{figure}
  \begin{minipage}[b]{0.5\linewidth}
    \centering \includegraphics[totalheight=2.5in]{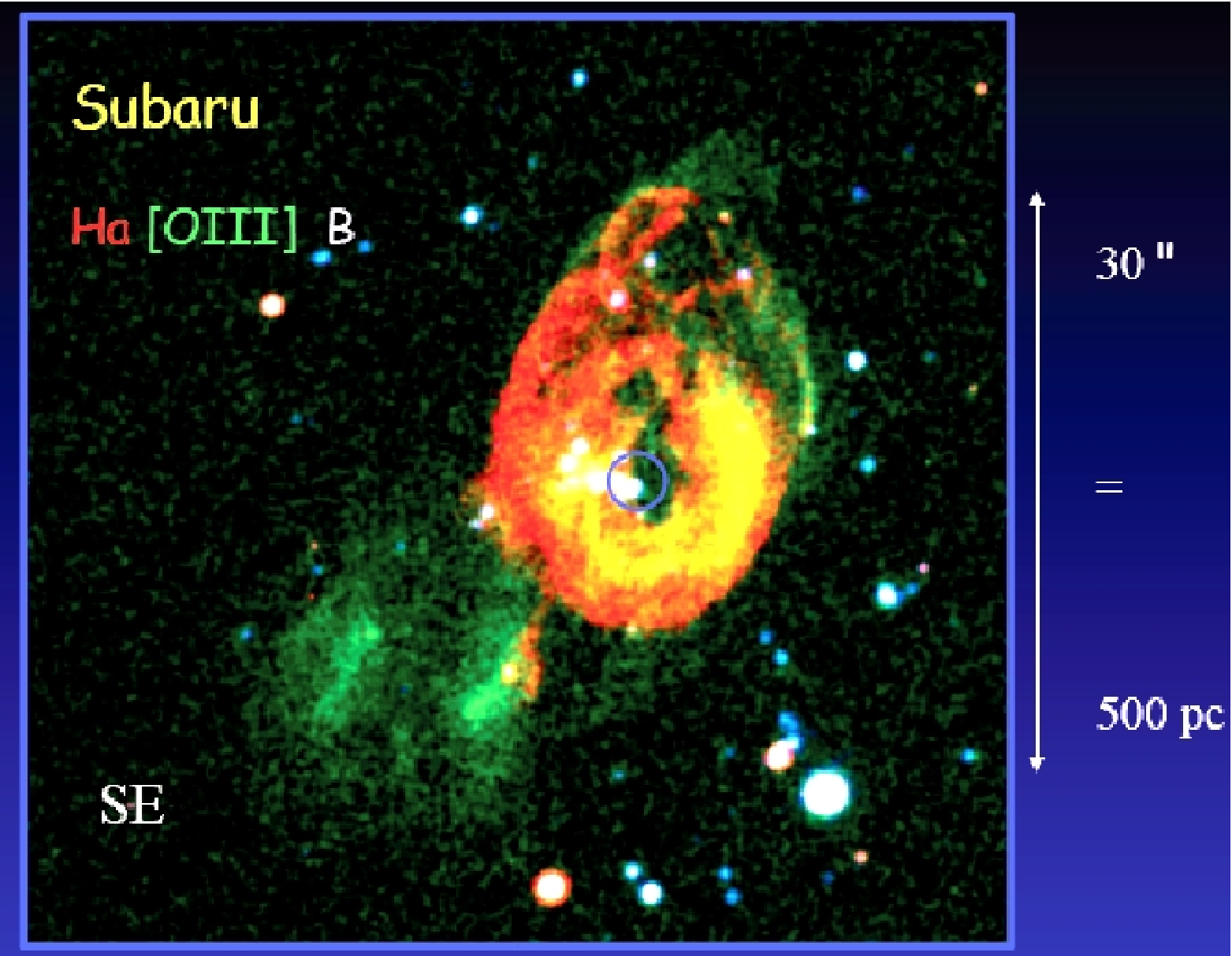} 
    \caption{Multi-colour image of the ULXB MH9-11 around
    Holmberg~IX X-1. Note the change of excitation [OIII]/H$\alpha$
    across the nebula and the green-coded OIII dominated emission to the
    SE indicating the presence of X-ray illuminated uncomplete shocks.}
  \end{minipage}
  \begin{minipage}[b]{0.5\linewidth}
    \centering \includegraphics[totalheight=2.5in]{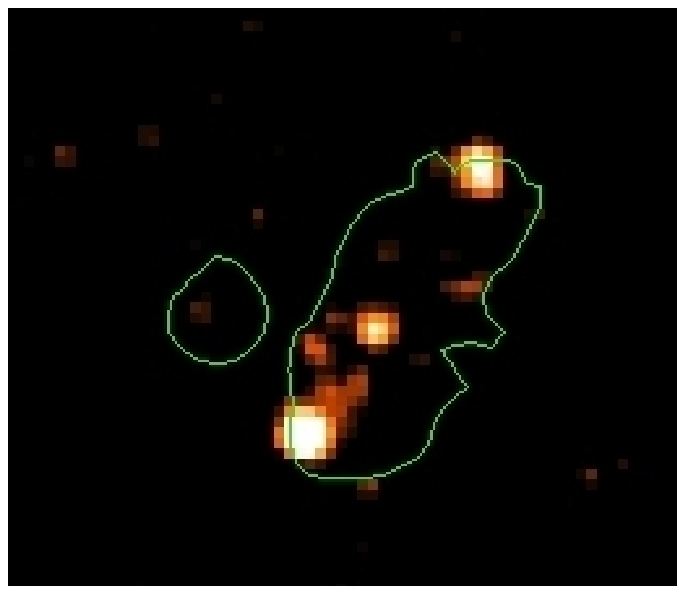}  
    \caption{The Chandra triple X-ray source in the nearby galaxy NGC~7793
    superimposed on the outer H$\alpha$ intensity contour 
    of the very large SNR candidate S26. The external sources 
    (beambags?) are $15\arcsec$ $\cong 250$ pc apart and match the 
    endpoints of the elongated bubble.}
  \end{minipage}
\end{figure}

\section{What ULX Bubbles can tell us about the Central Sources and their Jets}

An obvious question to ask is how ULX bubbles are being formed.  In
principle they could be superbubbles (SBs) blown by the combined
action of  many O stars and supernova remnants. However, two
considerations  argue against this interpretation:  known SBs don't
expand fast enough to create the fast shocks that are seen 
in the SNR-like spectra of ULXBs. And perhaps more convincingly, we point 
out that a mechanical energy input of \mbox{10$^{40}$ erg/s,} equivalent to 10-50
SN explosions within 10$^6$ years, requires a massive cluster of at
least $\sim10^6$ M$_{\odot}$. However, what we see in the case of
NGC1313 X-2 and Holmberg~IX X-1 are stellar associations with masses
below $10^4$ M$_{\odot}$ \citep{gr08}; in other words, the expected
mechanical power of the observed population falls short by two
orders of magnitude, or more.

Alternatively, there is the possibility of a rare energetic
hypernova injecting \mbox{$>10^{52}$ erg} into the ISM \citep{no03,rob03};
i.e. most probably the event that created the BH in the ULX system.
Although such a scenario has again recently been proposed by
\citet{lm07} to account for a large non-thermal radio bubble in the
Local Group galaxy IC~10, we think that the alternative scenario
of a ULX wind/jet blown bubble is more likely, not least for the
following reason: 

After the SN explosion and the formation of a BH in the binary
system, mass transfer from the companion commences only
after a certain time span in which the new elliptical orbit 
circularizes
by  tidal forces, and during which the star evolves and expands on
a nuclear  time scale (few 10$^6$ yrs) until it fills its Roche
lobe. The aforementioned ULXB MF16 in NGC~6946  is the
smallest, most rapidly expanding, and  therefore youngest object in
this group with a (Sedov) age of 2.5 10$^4$ yrs \citep{du00}.
Indeed, this period  would by far be too short for the onset of
mass transfer in this system. 
We therefore believe that MF 16 provides ample evidence that 
hypernov\ae\ are not the events that create ULX bubbles.
Instead, the view that ULXBs are being inflated by
quasi-relativistic winds or jets with mechanical power of some
10$^{39-40}$ erg/s (i.e. not unlike the famous SS433/W50  "beambag"
system \citep{beg80}) begins to materialize. Support for this
idea also comes from the discovery of a small (by ULXB standards)
$\sim5$ pc diameter radio and optical bubble around the famous black hole
X-ray binary Cyg X-1 \citep{gal05} which we now interpret as a
downscaled version (in jet power) of the bubbles described here, i.e. as
\mbox{a '$\mu$ULXB'.}

\section{More Tests with ULXBs}

Considering that extragalactic ultraluminous X-ray sources are
often  (we estimate in $>30\%$ of the cases) surrounded by large
ULXB  we might test the hypothesis of beamed emission by searching
for  (isotropically emitting) ULXB-like nebul\ae\ without
associated ULX. To be more specific: assuming a mild beaming factor
of  10 as often proposed \citep{king} we would expect, 
for each visible ULX, to see about ten such systems that 
are pointed away from
us. To that end, we have looked for large shocked ionized nebul\ae\
in our  VLT images of the galaxy NGC~1313 which harbours two
ultraluminous sources, X-1 and X-2, both of which are embedded in 
ULXBs \citep{pam02}.  Although we find one possible such case, we
certainly don't see  20 such nebul\ae\ without associated 
compact X-ray sources in  that galaxy.  Therefore, we conclude 
that ULXs emit largely isotropically, as previously shown  for 
Holmberg~II X-1 on the basis of optical Zanstra-method arguments.

Large shock-ionized bubbles without detectable X-ray point source
content may well represent systems that once contained ULX, but which
are no longer active; consider a typical 1 million years old standard
ULXB "beambag" of 400 pc diameter and 100 km/s expansion velocity which
is suddenly being deprived of wind/jet mechanical power input. What
would happen? The answer is that it  would still be visible for another
3$\times$10$^6$ years. Here we have assumed that the decreasing expansion
velocity must still be significantly supersonic (i.e. $\geq40$ km/s) in
order to form an observable radiative shock. 
In view of some ULXBs that had previously been listed in SNR 
catalogues of nearby galaxies -- sometimes as overlapping remnants 
that turned out to be part of one bigger structure \citep{pam03} -- we 
have extended our search by looking for point-like X-ray sources at 
the position of unusually large ($\geq100$ pc) SNRs. This exercise 
has already revealed some interesting results:

\begin{itemize}
 \item The well-known variable ULX NGC\,2403 X-1 which is located near 
the galactic nucleus lies in between the SNR candidates MFBL\,14 and 15
listed by \citet{metal97}. This has lead \citet{sp03} to effectively
discard a likely association. However, archival HST H$\alpha$ images show 
that the individual SNRs are part of a larger structure of
some 300 pc diameter, i.e., here is a new textbook ULXB.
 \item In their survey for SNR in the 7.0 Mpc distant spiral NGC~5885, 
\cite{mf97} pointed out the enormous dimension of their candidate \#1
(we measured $200\times300$ pc). It has the form of a regular 
elongated bubble, not unlike the ULXB around NGC~1313 X-2 \citep{pagm06}.
Unfortunately, an earlier proposal for XMM/Newton  observations of this
galaxy was turned down, but a recent public Chandra ACIS image now
reveals a bright point source with a luminosity of about 
\mbox{10$^{39}$ erg/s} right in the middle of the 'remnant': 
another beautiful example of a ULXB. 
\item Finally, we mention the Chandra X-ray counterpart of the
well-known  huge SNR candidate S26 in the Sculptor group galaxy NGC\,7793
\citep{bl97}, which is  bright at both radio and optical wavelengths
\citep{pan02}. Deep H$\alpha$ images show an elongated bubble
with some filamentary structure and large extent of 300 pc.
Overlaying the outer H$\alpha$ contour on a deep archival Chandra image 
reveals a very interesting configuration shown in Fig. 2: 
Here the X-ray source is resolved into a linear triplet, with the outer 
very soft sources being coincident with
the endpoints of the large axis of the optical bubble. Similarities
with X-ray images of SS433 and its X-ray lobes \citep{wat83} are striking, 
and we believe that this system is indeed the long-searched (for nearly 30
years) extragalactic analogue of this famous microquasar. We finally
mention that the observed X-ray output \mbox{(L$_x\sim10^{37}$ erg/s)} of the
central source is far from 'ultraluminous', a property it shares with
the X-ray weak SS433. Future work on S26 will certainly shed more light on the
controversial relations between SS433 and ULXs on the one hand, and
between W50 and ULXB beambags on the other hand.
\end{itemize}

\acknowledgements MWP wishes to thank the organizers for an inspiring 
conference and the editors for their patience with the delivery of this 
manuscript.

\end{document}